\documentclass{ptapap}
\usepackage{amsmath}
\usepackage{amssymb}

\usepackage{color}

\usepackage{soul}

\author{Rodolfo Smiljanic}[CAMK]
\affil[CAMK]{Nicolaus Copernicus Astronomical Center, Polish Academy of Sciences, Bartycka 18, 00--716 Warsaw, Poland}

\title{Near-UV Spectroscopy with the VLT}

\begin{document}

\maketitle

\begin{abstract}

The 39-meter European Extremely Large Telescope (E-ELT) is expected to have very low throughput in the blue part of the visible spectrum. Because of that, a blue-optimised spectrograph at the 8-meter Very Large Telescope could potentially be competitive against the E-ELT at wavelengths shorter than 400 nm. A concept study for such an instrument was concluded in 2012. This would be a high-throughput, medium resolution (R $\sim$ 20\,000) spectrograph, operating between 300 and 400 nm. It is currently expected that construction of this instrument will start in the next few years. In this contribution, I present a summary of the instrument concept and of some of the possible Galactic and extragalactic science cases that motivate such a spectrograph. 

\end{abstract}

\section{Introduction}\label{sec:intro}

Most future ground-based large telescopes, including the European Extremely Large Telescope (E-ELT), seem to focus their planned instrumentation on the red to near-infrared wavelengths. Therefore, it was recognised that an efficient spectrograph dedicated to the ground-visible near-UV wavelengths (300-400 nm) could become a unique capability of the VLT (Very Large Telescope). Indeed, a study of the MOSAIC multi-object spectrograph for the E-ELT recently concluded that a blue-optimised spectrograph at the VLT could be competitive against the E-ELT at wavelengths bluer than 400 nm \citep{2016SPIE.9908E..9JE}.

In this context, a study of such an instrument, at the time called CUBES (Cassegrain U-band Brazilian-ESO Spectrograph), was completed in 2012 \citep[see description in][]{2014Ap&SS.354..191B,2014SPIE.9147E..09B}. CUBES was envisioned to be a high-efficiency, medium-resolution spectrograph (R $\sim$ 20\,000) dedicated to the near-UV. 

Similar capabilities have been available before, even within ESO telescopes \citep[see, e.g.,][]{2014Ap&SS.354..121P}. Nevertheless, with improvements leading to higher efficiency, CUBES could enable access to samples 2-3 magnitudes deeper than currently possible with UVES \cite[UV-Visual Echelle Spectrograph,][]{2000SPIE.4008..534D}, the high-resolution spectrograph with near-UV capabilities currently available at the VLT. 

Even though, for various reasons, the project did not yet proceed to the construction stage, CUBES has always figured as part of the instrumentation plans for the VLT in the years after 2020 \citep[e.g.,][]{2013Msngr.154....2P,pasquini_luca_2019_3356250}. As of November 2019, ESO is preparing a call for a new phase A study for this near-UV spectrograph\footnote{\texttt{http://www.eso.org/public/industry/cp/CFP\_VLT\_UV\_Spectro\_PhaseA.html}}, giving new momentum to the project. In anticipation to this call, a group interested in forming a consortium to build the instrument was formed and the science case of the instrument was revisited and updated \citep[see][]{2018SPIE10702E..2EE,evans_chris_2019_3356264}.

In this contribution, I first summarise the preliminary instrument concept and some of the developments since the original phase A of 2012 (Section \ref{sec:cubes}). Afterwards, Sections \ref{sec:gal} and \ref{sec:extragal} summarise some of the science cases that are expected to be addressed with CUBES in the Galactic and extragalactic contexts. For more details, the interested reader is referred to the original phase A science case, available in the website of the UV astronomy workshop held at ESO in 2013\footnote{\texttt{http://www.eso.org/sci/meetings/2013/UVASTRO/1\_VLT-TRE-ESO-13800-5679\_i1\_CUBES\_SciCase.pdf}} and to the references given in this text.
   
\section{The Spectrograph}\label{sec:cubes}

An overview of the original instrument design is given in \citet{2014Ap&SS.354..191B}. A first study of the science cases concluded that the spectrograph should achieve at least a resolution of 20\,000 and wavelength coverage of 310-360 nm, with a goal of extending to 302-385 nm.

The spectrograph is envisioned to be a simple and straightforward instrument, minimising the number of parts and mechanisms to only those which are strictly needed. The main driver in this aim of simplicity is the desire to maximise the throughput and thus the efficiency of the spectrograph. For example, the instrument is expected to be mounted at one of the VLT Cassegrain foci, taking advantage of the higher throughput as compared to a Nasmyth focus. 

% The main driver for the resolution and wavelength coverage is the study of absorption lines in the spectra of Galactic metal-poor stars, 

To preserve efficiency, cross dispersers will not be used, instead the use of an image slicer is envisioned to achieve the needed resolution. The initial design included three slices of 0.375$''$ (but with a feasibility study still pending). In principle, the instrument would have an array of three 4k x 4k detectors. A study to understand if any important spectral feature would fall between the detector gaps is still needed. Without a cross disperser, there is the need to carefully balance the trade offs between resolution and wavelength range for a given size of the detector array.

The instrument design uses a transmission grating manufactured with the same technology as the grating developed for the \emph{Gaia} mission \citep{2012SPIE.8450E..2ZZ}. In one of the most important developments since the original 2012 study, a prototype grating for CUBES has been developed by the German Fraunhofer Institute for Applied Optics and Precision Engineering\footnote{\texttt{https://www.iof.fraunhofer.de/en.html}} with funding from a Brazilian grant. This prototype grating has 250mm $\times$ 130mm and three of them would be needed for CUBES. First tests of this prototype have shown that the diffraction efficiency is larger than 60 \% in the relevant spectral range, going up to almost 90\% at 320nm \citep{burmeister}.

\section{Galactic Science}\label{sec:gal}

Perhaps the main scientific motivation for the instrument is the study of the chemical composition of old metal-poor stars for the understanding of the formation and early evolution of the Milky Way. Such old stars are mostly distant and are thus faint, and therefore enabling access to a sample which is 2 to 3 magnitudes deeper than currently possible becomes of high interest for such studies. Nevertheless, it is expected that the instrument will open new parameter space for a broad range of topics, becoming of interest for a larger community.

For stellar studies, the spectral range of CUBES contains atomic lines of the light element Be, of heavy elements produced by the r- and s-processes, and molecular lines of CN, OH, and NH. For Be, in particular, this region contains the only lines of this element that are observable from the ground at 313.042 and 313.107 nm. 

Abundances of beryllium are interesting for a variety of applications. The science around Be abundances is a core part of the scientific motivation for the CUBES instrument \citep{2014Ap&SS.354...55S}. For studies of stellar evolution, Be abundances can help in the understanding of the physical mechanisms behind the transport of chemicals and angular momentum in low-mass stars \citep[e.g.,][]{2010A&A...510A..50S}. Moreover, because the only nucleosynthetic origin of Be is through cosmic-ray spallation, it has been suggested that, in the early Galaxy, Be abundances could be used as a cosmochronometer \citep{2005A&A...436L..57P}. Finally, in globular clusters, because there are no stellar sources of this element, Be abundances become a unique constraint of the dilution-pollution scenarios currently proposed to explain the chemical properties observed in these systems \citep{2004A&A...426..651P}. 

The near-UV wavelength range contains spectral lines of a large number of elements heavier than Fe \citep{2014Ap&SS.354...41S}. Recently, it has been shown that a kilonova, resulting from the merger of two neutron stars, can produce r-process elements \citep{2017Natur.551...67P}. Nevertheless, this is likely not the only nucleosynthetic channel producing these elements. CUBES will enable studies of a larger sample of metal-poor stars that can eventually shed new light on the production channels of these heavy elements. Moreover, on the topic of neutron star mergers themselves, such an instrument will allow the follow up of the spectral evolution of such events in the near-UV spectral range.

The near-UV region is also interesting for solar system science. For example, OH emission at 308 nm is a diagnostic of water sublimation in comets. An efficient instrument like CUBES can enable the study of faint comets, leading to the possibility of mapping the water distribution in the solar system \citep{snodgrass_colin_2019_3356322}.

CUBES is also foreseen to be important in studies of objects like planetary nebulae, symbiotic stars, X-ray binaries, and novae. Interested readers are referred to \citet{2014Ap&SS.354..191B} for a description of these possibilities.

\section{Extragalactic Science}\label{sec:extragal}

Depending on redshift, a number of lines from both the broad and narrow line regions of active galactic nuclei and from quasars can be present in the near-UV region \citep[see][for details]{2014Ap&SS.354..191B}. With CUBES, it will be possible to observe rest-frame far-UV features of galaxies in redshifts between 1.5 and 2.5. This is complimentary to what will be possible with the E-ELT MOSAIC instrument, as the same features can be detected in visible wavelengths for galaxies with redshift between 2.5 and 3.5 \citep{2018SPIE10702E..8RP}.

With CUBES, it will be possible to observe the Lyman limit of damped Lyman-$\alpha$ systems down to redshift $\sim$ 2.4. It might be possible to detect uncontaminated metal lines (of, e.g., D, O I, N I, S II) in the spectra of these objects. Considering that damped Lyman-$\alpha$ systems are likely galaxies at an early stage of their evolution, studies of these objects can contribute to the understanding of the early chemical enrichment of the Universe \citep{2014Ap&SS.354...75M}.

Other possible science cases include the search for molecular hydrogen in the interstellar medium of high-redshift galaxies, studies of the cosmic UV background to understand the sources that drive reionisation, and investigations of the interaction between galaxies and the inter-galactic medium. Interested readers are referred to \citet{2014Ap&SS.354..191B} and \citet{2018SPIE10702E..2EE} for a description of these cases.

\acknowledgements{R.S. acknowledges support by the Polish National Science Centre through project 2018/31/B/ST9/01469.}

\bibliographystyle{ptapap}
\bibliography{smiljanic}

\begin{thebibliography}{20}
\providecommand{\natexlab}[1]{#1}
\providecommand{\url}[1]{\texttt{#1}}
\providecommand{\urlprefix}{URL }
\providecommand{\eprint}[2][]{\url{#2}}

\bibitem[{{Barbuy} et~al.(2014)}]{2014Ap&SS.354..191B}
{Barbuy}, B., et~al., \emph{\apss} \textbf{354}, 191 (2014)

\bibitem[{{Bristow} et~al.(2014)}]{2014SPIE.9147E..09B}
{Bristow}, P., et~al., in Ground-based and Airborne Instrumentation for
  Astronomy V, \emph{\procspie}, volume 9147, 914709 (2014)

\bibitem[{{Burmeister} et~al.(2018)}]{burmeister}
{Burmeister}, F., et~al., in Advances in Optical and Mechanical Technologies
  for Telescopes and Instrumentation III, Paper 10706-74 (2018)

\bibitem[{{Dekker} et~al.(2000)}]{2000SPIE.4008..534D}
{Dekker}, H., et~al., in M.~{Iye}, A.~F. {Moorwood} (eds.) Optical and IR
  Telescope Instrumentation and Detectors, \emph{Society of Photo-Optical
  Instrumentation Engineers (SPIE) Conference Series}, volume 4008, 534--545
  (2000)

\bibitem[{Evans(2019)}]{evans_chris_2019_3356264}
Evans, C.~J. (2019), \urlprefix\url{https://doi.org/10.5281/zenodo.3356264}

\bibitem[{{Evans} et~al.(2016)}]{2016SPIE.9908E..9JE}
{Evans}, C.~J., et~al., in Ground-based and Airborne Instrumentation for
  Astronomy VI, \emph{\procspie}, volume 9908, 99089J (2016)

\bibitem[{{Evans} et~al.(2018)}]{2018SPIE10702E..2EE}
{Evans}, C.~J., et~al., in Ground-based and Airborne Instrumentation for
  Astronomy VII, \emph{Society of Photo-Optical Instrumentation Engineers
  (SPIE) Conference Series}, volume 10702, 107022E (2018)

\bibitem[{{Molaro}(2014)}]{2014Ap&SS.354...75M}
{Molaro}, P., \emph{\apss} \textbf{354}, 1, 75 (2014)

\bibitem[{{Pasquini}(2014)}]{2014Ap&SS.354..121P}
{Pasquini}, L., \emph{\apss} \textbf{354}, 1, 121 (2014)

\bibitem[{Pasquini(2019)}]{pasquini_luca_2019_3356250}
Pasquini, L. (2019), \urlprefix\url{https://doi.org/10.5281/zenodo.3356250}

\bibitem[{{Pasquini} et~al.(2013){Pasquini}, {Casali}, \&
  {Russell}}]{2013Msngr.154....2P}
{Pasquini}, L., {Casali}, M., {Russell}, A., \emph{The Messenger} \textbf{154},
  2 (2013)

\bibitem[{{Pasquini} et~al.(2004)}]{2004A&A...426..651P}
{Pasquini}, L., et~al., \emph{\aap} \textbf{426}, 651 (2004)

\bibitem[{{Pasquini} et~al.(2005)}]{2005A&A...436L..57P}
{Pasquini}, L., et~al., \emph{\aap} \textbf{436}, L57 (2005)

\bibitem[{{Pian} et~al.(2017)}]{2017Natur.551...67P}
{Pian}, E., et~al., \emph{\nat} \textbf{551}, 67 (2017)

\bibitem[{{Puech} et~al.(2018)}]{2018SPIE10702E..8RP}
{Puech}, M., et~al., in \procspie, \emph{Society of Photo-Optical
  Instrumentation Engineers (SPIE) Conference Series}, volume 10702, 107028R
  (2018)

\bibitem[{{Siqueira-Mello} \& {Barbuy}(2014)}]{2014Ap&SS.354...41S}
{Siqueira-Mello}, C., {Barbuy}, B., \emph{\apss} \textbf{354}, 1, 41 (2014)

\bibitem[{{Smiljanic}(2014)}]{2014Ap&SS.354...55S}
{Smiljanic}, R., \emph{\apss} \textbf{354}, 55 (2014)

\bibitem[{{Smiljanic} et~al.(2010){Smiljanic}, {Pasquini}, {Charbonnel}, \&
  {Lagarde}}]{2010A&A...510A..50S}
{Smiljanic}, R., {Pasquini}, L., {Charbonnel}, C., {Lagarde}, N., \emph{\aap}
  \textbf{510}, A50 (2010)

\bibitem[{Snodgrass(2019)}]{snodgrass_colin_2019_3356322}
Snodgrass, C. (2019), \urlprefix\url{https://doi.org/10.5281/zenodo.3356322}

\bibitem[{{Zeitner} et~al.(2012){Zeitner}, {Fuchs}, \&
  {Kley}}]{2012SPIE.8450E..2ZZ}
{Zeitner}, U.~D., {Fuchs}, F., {Kley}, E.~B., in Modern Technologies in Space-
  and Ground-based Telescopes and Instrumentation II. Proceedings of the SPIE,
  Volume 8450, article id. 84502Z, 7 pp. (2012)., \emph{Society of
  Photo-Optical Instrumentation Engineers (SPIE) Conference Series}, volume
  8450, 84502Z (2012)

\end{thebibliography}

\end{document}